\documentclass[a4paper,11pt]{article}

\usepackage{jheppub} 
\usepackage{subfigure}
\usepackage{subfigure}
\usepackage{ascmac}
\usepackage{amsthm}
\usepackage{amsmath}
\usepackage{tikz}
\usetikzlibrary{arrows.meta}
\usetikzlibrary{patterns}
\usetikzlibrary{decorations.pathmorphing}
\usetikzlibrary{decorations.text}
\tikzset{>={Latex[width=2mm,length=2mm]}}

\preprint{OU-HET-1095}

\title{\boldmath 
Bulk reconstruction of metrics 
inside black holes \\
by complexity
}

\author{Koji Hashimoto,}
\affiliation{Department of Physics, Osaka University,\\1-1 Machikaneyama, Toyonaka, Osaka 560-0043, Japan}
\author{Ryota Watanabe}

\emailAdd{koji@phys.sci.osaka-u.ac.jp}
\emailAdd{watanabe@het.phys.sci.osaka-u.ac.jp}

\abstract{
We provide a formula to reconstruct bulk spacetime metrics inside black holes by the time dependence of complexity in the dual quantum field theory, based on the complexity=volume (CV) conjecture in the holographic duality.
}

\begin{document}

\maketitle
\flushbottom

\section{Introduction}
\label{sec:intro}

It is expected that the holographic principle and
the AdS/CFT correspondence \cite{Maldacena:1997re,Gubser:1998bc,Witten:1998qj}
will serve as a solution to the information loss problem of black holes.
For that program to be complete, it is indispensable to find out 
the way how the information inside black holes is encoded in
the dual quantum field theories.
Can we construct the spacetime metric inside a black hole horizon by a given data of the dual 
quantum field theory?

Bulk reconstruction is in general an inverse problem in holography, thus is difficult. Nevertheless,
various methods have been developed to reconstruct bulk spacetime metrics 
by the data of dual quantum field theories (QFTs)\footnote{Here we focus on only the bulk reconstruction of spacetime metrics. Reconstruction of operators in the bulk AdS spacetimes \cite{Dong:2016eik}
have wider references.}. 
Successful methods include
the holographic renormalization \cite{deHaro:2000vlm}, the reconstruction using bulk geodesics and light cones 
\cite{Hammersley:2006cp,Hubeny:2006yu,Engelhardt:2016wgb,Engelhardt:2016crc,Roy:2018ehv,Burda:2018rpb,Hernandez-Cuenca:2020ppu}, the reconstruction \cite{Hammersley:2007ab,Bilson:2008ab,Bilson:2010ff,Hubeny:2012ry,Balasubramanian:2013rqa,Balasubramanian:2013lsa,Myers:2014jia,Czech:2014wka,Czech:2014ppa,Gentle:2015cfp,Saha:2018jjb,Bao:2019bib,Cao:2020uvb,Jokela:2020auu} using holographic entanglement entropy \cite{Ryu:2006bv,Ryu:2006ef}
\footnote{Related methods include the one \cite{Swingle:2012wq,Bao:2018pvs,Milsted:2018san,Bao:2019fpq} using tensor networks \cite{Swingle:2009bg,Pastawski:2015qua} through the entanglement properties.}, the inversion formula \cite{Hashimoto:2020mrx} of the holographic Wilson loops \cite{Maldacena:1998im,Rey:1998ik,Brandhuber:1998bs,Rey:1998bq},
and the machine learning holography \cite{Hashimoto:2018ftp,Hashimoto:2018bnb,Tan:2019czc,Yan:2020wcd,Akutagawa:2020yeo,Hashimoto:2020jug}\footnote{The holographic bulk spacetime is identified with neural networks 
\cite{You:2017guh,Hashimoto:2018ftp,Hashimoto:2018bnb,Hu:2019nea,Hashimoto:2019bih,Tan:2019czc,Yan:2020wcd,Akutagawa:2020yeo,Hashimoto:2020jug,Song:2020agw}, and the spacetimes are emergent.
See \cite{Ruehle:2020jrk} for a review of data science approach to string theory, and also see
\cite{Carleo:2019ptp} for applications of machine learning to material sciences.}.
However, all of these methods do not reconstruct the black hole interior, because the
dual thermal 
QFT data is associated with the asymptotic boundary of the bulk spacetime, 
and the bulk classical 
gravity does not extract information inside the horizon.

The QFT quantity to probe the inside of the bulk black holes in holography is the quantum computational complexity, proposed by Susskind \cite{Susskind:2014rva, Susskind:2014moa, Susskind:2016tae}.
The conjecture originates from the idea using two-sided extended black hole geometry interpreted as a doubled QFT with thermo-field double states \cite{Maldacena:2001kr} 
and the use of the Ryu-Takayanagi (RT) surface connecting the left and right boundaries of the geometry \cite{Hartman:2013qma}.
There are two versions of the complexity conjecture, 
one is the ``complexity = volume" (CV) conjecture \cite{Stanford:2014jda, Susskind:2014jwa}, 
and the other is the ``complexity = action" (CA) conjecture \cite{Brown:2015bva, Brown:2015lvg}. 
Both conjectures can probe the interior of the horizons of the maximally extended black hole spacetimes.
Therefore, the question of how to reconstruct even the interior geometry of the black hole 
can be cast into a problem of unifying the bulk reconstruction techniques and the complexity conjecture.

In this paper, based on the CV conjecture,
we provide a formula to reconstruct the spacetime metric of the black hole interior
by the time dependence of the complexity ${\cal C}(t)$ of the dual QFT.
The CV conjecture \cite{Stanford:2014jda, Susskind:2014jwa}
states that the quantum computational complexity of a thermo-field double state at a time in the dual QFT is equal to the volume of an extremal surface in the bulk penetrating the horizon and anchored to the boundary at the QFT time. 
Among CV and CA conjectures, we employ the CV conjecture, because in the bulk side it is about extremal surfaces, and therefore it allows a similarity to the RT surface and the  holographic Wilson loops.
For the latter quantities there exist the particular 
bulk reconstruction techniques for the spacetime geometry
\cite{Bilson:2008ab,Bilson:2010ff,Hashimoto:2020mrx}
to invert the area of extremal surfaces back to the metric. By generalizing the technique\footnote{
Note that the interesting interplay found for the RT surface and the holographic Wilson loop in thermalization and confinement \cite{Balasubramanian:2010ce,Balasubramanian:2011ur,Kol:2014nqa,Bao:2019hwq}
is about the outside of the black holes.}
to the case of the extremal surfaces penetrating the black hole horizons, 
we will be able to reconstruct the metric of the black hole interior by
the complexity of the dual QFT.

We focus on generic black hole spacetimes which are static, spherically symmetric and 
asymptotically AdS.
They have the metric
\begin{align}
ds^2 = -f(r) dt^2 + g(r) dr^2 + r^2 d\Sigma^2 \, .
\label{metfg}
\end{align}
Let us make technical remarks on our reconstruction. 
First, because the metric includes two functional degrees of freedom 
$f(r)$ and $g(r)$, 
the complexity ${\cal C}(t)$ can determine only one functional degree of freedom
in the metric.
Thus, when $g(r)$ and $f(r)$ are related, the complexity ${\cal C}(t)$ can reconstruct the metric.
On the other hand, when they are independent, 
we need another QFT data, and we employ the Hartman-Maldacena entanglement entropy $S_{\rm HM}(t)$
\cite{Hartman:2013qma} together with the complexity ${\cal C}(t)$. 
Second, in this paper we are interested in only the interior of the black hole, so 
we assume that the metric outside the horizon is known by some methods listed above.

This paper is organized as follows. 
In Sec.~\ref{sec:2}, 
we provide our formula for reconstructing bulk metrics inside the black hole horizons 
by the time dependence of the complexity ${\cal C}(t)$ in the dual QFT, when $g(r)=1/f(r)$.
In Sec.~\ref{sec:3}, 
we provide our formula for the general case of unknown $f(r)$ and $g(r)$, 
by the complexity ${\cal C}(t)$ and the Hartman-Maldacena entanglement entropy $S_{\rm HM}(t)$.
In Sec.~\ref{sec:4}, 
to illustrate the validity of our formulas, we verify the reconstruction of the interior of
the BTZ black hole.
Sec.~\ref{sec:5} is for our summary and discussions.
App.~\ref{app:g(f) case} treats a special case when $g(r)$ is known as a function of $f(r)$ implicitly, $g=g(f)$.

\vspace{20mm}


\section{Reconstructing metrics inside horizons by complexity}
\label{sec:2}

In this section, we provide the formula to reconstruct the interior of a black hole by the 
complexity in the dual QFT, based on the complexity=volume (CV) conjecture \cite{Stanford:2014jda,Susskind:2014moa}
\begin{equation}
\label{eq:CV}
	{\cal C}(t)=\frac{V(t)}{G_{\rm N} L}\, .
\end{equation}
Here, $V(t)$ is the volume of the extremal surface defined in the two-sided eternal AdS black holes.
The surface is anchored at time $t$ at the two boundaries of the spacetime. 
$G_{\rm N}$ is the Newton's constant and $L$ is the AdS radius.

\subsection{Review: calculation of the volume}
\label{subsec:2-1}

We here first summarize the known results on the calculation of the volume $V(t)$ 
in the gravity side, for a given metric.
Following \cite{Carmi:2017jqz}, 
in this section we consider the metric of the form\footnote{Our notations follow those in \cite{Carmi:2017jqz}.}
\begin{equation}
\label{eq:f metric}
	ds^2=-f(r)dt^2+\frac{dr^2}{f(r)}+r^2d\Sigma_{k,d-1}^2 \, .
\end{equation}
The dimension of the bulk spacetime is $d+1$, and $k=\{+1,0,-1\}$ indicates the curvature of the $(d-1)$-dimensional line element $d\Sigma_{k,d-1}^2$. 
The geometry is asymptotically AdS,
$f(r)\to\frac{r^2}{L^2}$ as $r\to \infty$.

The function $f(r)$ may be determined as a solution of Einstein equations. However, for our purpose
of the reconstruction, we keep $f(r)$ to be a generic function. The metric \eqref{eq:f metric} corresponds
to having $g(r)=1/f(r)$ in \eqref{metfg}.
For ablack holes with a nonzero value of the temperature,
in the vicinity of the horizon the function $f(r)$ is linear, and at the horizon $r=r_{\rm h}$
we have $f(r=r_{\rm h})=0$.

Of our interest is the codimension-one extremal surface which penetrates the horizon and connects the left and right boundaries in the maximally extended geometry. Particularly, we are interested in the case where the boundary times $t_{\rm L}$ and $t_{\rm R}$ are the same, $t_{\rm L}=t_{\rm R}$, so that the surface is left-right symmetric\footnote{We define $t_L$ and $t_R$ in such a way that both of them run from the past to the future.}. As in \cite{Carmi:2017jqz}, we can express the volume V of the extremal surface and the boundary time $t_{\rm R}$ in terms of the metric function $f(r)$ and the minimal radius $r_{\rm min}<r_{\rm h}$ of the surface\footnote{In Sec. \ref{subsec:volume}, we will explicitly derive them for a more general metric. See \eqref{eq:extremal volume} and \eqref{eq:t_R}.},
\begin{equation}
\label{eq:sec 2 volume}
	V=2\Omega_{k,d-1}\int_{r_{\rm min}}^{r_{\rm max}} dr~\frac{r^{2(d-1)}}{\sqrt{F(r)-F(r_{\rm min})}}\, ,
\end{equation}
\begin{equation}
\label{eq:sec 2 t_R}
	t_{\rm R}=\int_{r_{\rm min}}^{r_{\rm max}} dr~\frac{r^{2(d-1)}}{F(r)}\frac{-\sqrt{-F(r_{\rm min})}}{\sqrt{F(r)-F(r_{\rm min})}}\, ,
\end{equation}
where $F(r)\equiv f(r)r^{2(d-1)}$, and $r_{\rm max}$ is a UV cutoff\footnote{
The integral \eqref{eq:sec 2 volume} is divergent for $r_{\rm max}\to\infty$ while that
in \eqref{eq:sec 2 t_R} is convergent.}.
There are two important comments on these equations: 
First, since $r_{\rm min}$ is at the interior of the black hole, $F(r_{\rm min})=f(r_{\rm min})r_{\rm min}^{2(d-1)}$ is negative. Second, 
although the integrand of \eqref{eq:sec 2 t_R} blows up near the horizon $r\sim r_{\rm h}$, we can define the integral in terms of the Cauchy principal value.

In the time evolution, the surface gets stretched and the volume grows. 
The growth rate $dV/dt_{\rm R}$ can be found as
\begin{equation}
\label{eq:sec 2 dV/dt_R}
	\frac{dV}{dt_{\rm R}} = 2\Omega_{k,d-1} \sqrt{-F(r_{\rm min})}\, .
\end{equation}
At late times, this asymptotically approaches the constant value
\begin{equation}
\label{eq:sec 2 late time}
	\lim_{t_{\rm R}\to\infty}\frac{dV}{dt_{\rm R}}=2\Omega_{k,d-1}\sqrt{-F(r_{\rm f})}\, .
\end{equation}
Here, $r=r_{\rm f}$ is the radius at which $\sqrt{-F(r)}$ is maximal\footnote{
When there are several extremal radii of $\sqrt{-F(r)}$, this 
$r=r_{\rm f}$ is the one closest to the horizon. 
In other words, we do not consider a sudden jump in the complexity in its time evolution.}. As $t_{\rm R}\to\infty$, the extremal surface asymptotically approaches this $r=r_{\rm f}$ surface (called a ``nice" slice), 
which in turn means that $r_{\rm min}$ is always greater than $r_{\rm f}$.

If we knew the functional form of the metric $F(r)$, then eliminating $r_{\rm min}$ from
\eqref{eq:sec 2 t_R} and \eqref{eq:sec 2 dV/dt_R} would determine the complexity growth
$\dot{\cal C}(t_{\rm R}) = \frac{dV}{dt_{\rm R}}(t_{\rm R})/G_{\rm N}L$.
Our interest in this paper is the opposite, {\it i.e.}~the case when the metric $F(r)$ is unknown: can we reconstruct
the function $F(r)$ by a given function $\frac{dV}{dt_{\rm R}}(t_{\rm R})$?
The answer is yes, and in the next subsection we will provide the formula for the reconstruction.

\clearpage

\subsection{Reconstruction formula}
\label{subsec:reconstruction of f}

Let us provide the reconstruction formula of the metric inside horizons by complexity.
\begin{screen}
Assume that
\begin{itemize}
\item the time derivative of the complexity $\dot{\cal C}(t)$ in the dual QFT is given.
\\[-8mm]
\item $\dot{\cal C}(t)$ is monotonic in time.
\\[-8mm]
\item the metric $f(r)$ of \eqref{eq:f metric} outside the horizon $(r \geq r_{\rm h})$ is known.
\end{itemize}
Then, solve
\begin{equation}
\label{eq:sec 2 dC/dt_R}
	\frac{d{\cal C}}{dt_{\rm R}}(t_{\rm R})=\frac{2\Omega_{k,d-1}}{G_{\rm N}L}\sqrt{-F_{\rm min}}
\end{equation}
to get $t_{\rm R}$ as a function of $F_{\rm min}$. In addition, outside of the horizon, invert $F=F(r)\equiv f(r)r^{2(d-1)}$ to find $r=r(F)$. Substitute them into
\begin{equation}
\label{eq:Q}
	Q(F_{\rm min}) \equiv \frac{t_{\rm R}(F_{\rm min})}{\sqrt{-F_{\rm min}}} + \int_{\epsilon}^\infty dF~\frac{dr}{dF}\frac{r(F)^{2(d-1)}}{F}\frac{1}{\sqrt{F-F_{\rm min}}}\, ,
\end{equation}
with a positive infinitesimal parameter $\epsilon$. For $F\leq-\epsilon$, calculate
\begin{equation}
\label{eq:sec 2 y}
	y(F) \equiv \frac{1}{\pi}\frac{d}{dF}\int_F^{-\epsilon}dU~\frac{Q(U)}{\sqrt{U-F}}\, ,
\end{equation}
and integrate the differential equation
\begin{equation}
\label{eq:sec 2 diff eq}
	\frac{dr}{dF}\frac{r(F)^{2(d-1)}}{F} = y(F)
\end{equation}
to find $r=r(F)$. Finally, invert $r(F)$ to find $F(r)$, and the metric inside the black hole is given
 as $f(r)=F(r)r^{-2(d-1)}$.
\end{screen}

\begin{proof}
First, \eqref{eq:sec 2 dC/dt_R} is nothing but \eqref{eq:CV} and \eqref{eq:sec 2 dV/dt_R} with $F_{\rm min}\equiv F(r_{\rm min})<0$. 
Next, we consider rewriting \eqref{eq:sec 2 t_R}. 
Assuming that $F(r)$ outside the horizon is monotonic in $r$, $F=F(r)$ can be inverted to a single-valued function $r=r(F)$. 
%
%
Then we can transform the integration variable of \eqref{eq:sec 2 t_R} from $r$ to $F$:
\begin{equation}
\label{eq:proof t_R}
	t_{\rm R}(F_{\rm min}) = \int_{F_{\rm min}}^{\infty}dF~\frac{dr}{dF}\frac{r(F)^{2(d-1)}}{F}\frac{-\sqrt{-F_{\rm min}}}{\sqrt{F-F_{\rm min}}}\, .
\end{equation}
To make the expression simpler, we define the function
\begin{equation}
	y(F) \equiv \frac{dr}{dF}\frac{r(F)^{2(d-1)}}{F}\, .
\end{equation}
As pointed out after \eqref{eq:sec 2 t_R}, we interpret \eqref{eq:proof t_R} as the Cauchy principal value. Introducing an infinitesimally small parameter $\epsilon>0$, we rewrite \eqref{eq:proof t_R} as
\begin{equation}
	-\frac{t_{\rm R}(F_{\rm min})}{\sqrt{-F_{\rm min}}}=\left(\int_{F_{\rm min}}^{-\epsilon}+\int_{-\epsilon}^{\epsilon}+\int_{\epsilon}^{\infty}\right)dF~\frac{y(F)}{\sqrt{F-F_{\rm min}}}\, .
\end{equation}
If we take $\epsilon$ small enough, the second integral $\displaystyle \int_{-\epsilon}^{\epsilon}$ cancels by itself, and we obtain
\begin{equation}
	-\frac{t_{\rm R}(F_{\rm min})}{\sqrt{-F_{\rm min}}}=\left(\int_{F_{\rm min}}^{-\epsilon}+\int_{\epsilon}^{\infty}\right)dF~\frac{y(F)}{\sqrt{F-F_{\rm min}}}\, .
\end{equation}
We rewrite this as
\begin{equation}
	\int_{F_{\rm min}}^{-\epsilon}dF~\frac{y(F)}{\sqrt{F-F_{\rm min}}} = -\frac{t_{\rm R}(F_{\rm min})}{\sqrt{-F_{\rm min}}}-\int_{\epsilon}^{\infty}dF~\frac{y(F)}{\sqrt{F-F_{\rm min}}} \equiv Q(F_{\rm min})\, .
\end{equation}
This is nothing but \eqref{eq:Q}.
Since we do not know the metric inside the black hole, the left-hand side is unknown. On the other hand, the right-hand side can be calculated if we know the complexity and the metric outside the horizon. As can be easily checked, the solution of  this integral equation is
\begin{equation}
	y(F) = \frac{1}{\pi}\frac{d}{dF}\int_F^{-\epsilon}dU~\frac{Q(U)}{\sqrt{U-F}}\, .
\end{equation}
\begin{figure}[t]
	\centering
		\begin{tikzpicture}
			\draw[black, very thick, ->] (0,0) -- (4,0) node[right] {$r$};
			\draw[black, very thick, ->] (0,-2) -- (0,2) node[left] {$F$};
			\draw (0,0) node[left]{$O$} cos (0.8,-0.8) sin (1.6,-1.6);
			\draw (1.6,-1.6) parabola (3,2);
			\draw[black, dashed] (1.6,-1.6) -- (1.6,0) node[above]{$r_{\rm f}$};
			\node at (2.8,-0.3) {$r_{\rm h}$};
		\end{tikzpicture}
	\caption{The typical $r$ dependence of $F(r)=f(r)r^{2(d-1)}$ for $d>2$.}
	\label{fig:shape of F(r)}
\end{figure}
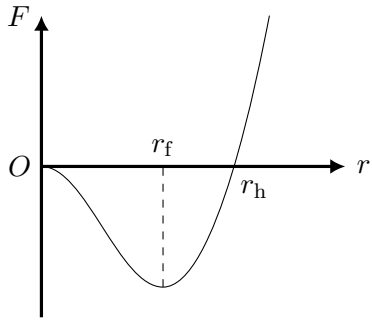
This provides the metric inside the horizon.
\end{proof}

\vspace{5mm}

Several comments on the formula are in order. In the proof above, first, 
we need to know 
the metric outside the horizon. In Sec.~\ref{sec:5} we discuss 
it is able to be reconstructed by the methods listed in 
Sec.~\ref{sec:intro} using some QFT data. 
Second, in the proof, we have assumed\footnote{
It is worthwhile to mention that the formula can be generalized such that the monotonicity of $F(r)$ outside the horizon is not necessary. The second term in \eqref{eq:Q} can be written as an integral over $r$
rather than the integral over $F$. So even with $F(r)$ which is not monotonic outside  the horizon,
the evaluation of $Q$ in \eqref{eq:Q} is possible. The boundary of the integration region 
$F=\epsilon$ needs to be rephrased in terms of $r$ by the known functional form of $F(r)$.} that the function $F(r)$ outside the horizon is monotonic in $r$.
The reconstructed $F(r)$ inside the horizon is also monotonic, which follows from the construction method. 
For example, in the AdS Schwarzschild geometry, 
$F(r)$ behaves like Fig.~\ref{fig:shape of F(r)}, and 
in the region $r_{\rm h}>r\geq r_{\rm f}$ it is monotonic.
%
Our reconstruction builds the spacetime of the region $r_{\rm h}>r\geq r_{\rm f}$, 
not the whole interior of the black hole. 
In particular, in general the reconstruction of
the black hole singularity is not possible, which is physically sensible.

\clearpage

\section{General case}
\label{sec:3}

In the previous section, we have demonstrated 
the reconstruction of the metric by the complexity 
assuming that the metric takes the particular form \eqref{eq:f metric}. 
In general the static and spherically symmetric black hole metric is of the form
\begin{equation}
\label{eq:f g indep case}
	ds^2=-f(r)dt^2 + g(r)dr^2 + r^2d\Sigma_{k,d-1}^2,
\end{equation}
where $f(r)$ and $g(r)$ are independent. 
In this section, we provide a reconstruction formula to determine both $f(r)$ and $g(r)$.\footnote{
See App.~\ref{app:g(f) case} for the reconstruction formula for the case when the relation between $f$ and $g$ is known,
$g=g(f)$.}

As is described in the introduction, the reconstruction in the previous section 
with \eqref{eq:f metric} is possible
since the time dependence of the complexity provided just enough degree of freedom 
to reconstruct a single unknown metric function $f(r)$. 
Therefore, 
for the reconstruction of both $f(r)$ and $g(r)$, 
we need another functional information in addition to the complexity. As such a quantity, we use the Hartman-Maldacena entanglement entropy \cite{Hartman:2013qma}.

\vspace{10mm}
\subsection{Preparations}
\label{subsec:volume}

To derive our reconstruction formula, in this subsection we first formulate the CV conjecture and
the Hartman-Maldacena entanglement entropy in generic black hole spacetimes \eqref{eq:f g indep case}.

\vspace{5mm}
\subsubsection{Volume of extremal surfaces in generic spacetimes}
\label{sec:3.1.1}

We express the volume of the extremal codimension-one surface using $f(r)$ and $g(r)$. This can be worked out by slightly generalizing the calculations in \cite{Carmi:2017jqz}. Again, let $r=r_{\rm h}$ be the position of the horizon. For black holes with a nonzero value of the temperature,
$f(r)$ and $1/g(r)$ are linear near the horizon and vanish at the horizon.
The spacetime is asymptotically AdS with radius $L$,
$f(r)\to\frac{r^2}{L^2}$, $g(r)\to\frac{L^2}{r^2}$ as $r\to \infty$.
We introduce the tortoise coordinate as
\begin{equation}
\label{eq:tortoise when f g indep}
	dr^*\equiv \frac{\sqrt{f(r)g(r)}}{f(r)}dr\,,
\end{equation}
and then define the incoming Eddington-Finkelstein coordinate $v\equiv t+r^*$. In terms of them, the metric \eqref{eq:f g indep case} is written as
\begin{equation}
\label{eq:f g indep metric in EF}
	ds^2=-f(r)dv^2+2\sqrt{f(r)g(r)}dvdr+r^2d\Sigma_{k,d-1}^2\, .
\end{equation}
Now, let us consider the codimension-one surface which penetrates the horizon and connects the left and right boundaries in the maximally extended geometry. We focus on the symmetric configuration, $t_{\rm L}=t_{\rm R}$. The profile of the surface is determined by $v(\lambda)$ and $r(\lambda)$, where $\lambda$ is a coordinate intrinsic to the surface. Then, the induced metric becomes
\begin{equation}
\label{eq:induced metric}
	ds^2=\left( -f(r)\dot{v}^2+2\sqrt{f(r)g(r)}\dot{v}\dot{r} \right)d\lambda^2+r^2d\Sigma_{k,d-1}^2\, ,
\end{equation}
where the dots represent $d/d\lambda$. We assume that, when the surface volume is extremized, $r(\lambda)$ is a strictly monotonically increasing function, $\dot{r}>0$. We set $\lambda=0$ at the left-right symmetric point on the surface, where the radius of the surface $r(\lambda)$ takes the minimum value, $r(\lambda=0)=r_{\rm min}<r_{\rm h}$. Smoothness requires that $\dot{r}(\lambda=0)=0$. Note that $r_{\rm min}$ varies depending on the boundary time $t_{\rm R}$. The volume of the surface is
\begin{align}
	V
	&=2\int d\lambda d\Sigma_{k,d-1}~\sqrt{\left( -f(r)\dot{v}^2+2\sqrt{f(r)g(r)}\dot{v}\dot{r} \right) r^{2(d-1)}} \notag \\
	&=2\Omega_{k,d-1} \int_{0}^{\lambda_{\rm max}} d\lambda \sqrt{-F(r)\dot{v}^2+2\sqrt{F(r)G(r)}\dot{v}\dot{r}}\, , \label{eq:volume}
\end{align}
where $F(r) \equiv f(r)r^{2(d-1)},\,G(r) \equiv g(r)r^{2(d-1)}$ and $\lambda_{\rm max}$ corresponds to the cutoff $r=r_{\rm max}$ necessary for the volume to be well-defined. 

Since we want to apply the CV conjecture, we need to extremize the volume while the boundary of the surface is fixed. To work it out, instead of solving the equations of motion for $v(\lambda)$ and $r(\lambda)$, we make use of a conserved quantity and the reparametrization invariance. Since the integrand of \eqref{eq:volume} is written only with $\dot{v}$ and does not contain $v$, 
when the volume is extremal,
\begin{equation}
\label{eq:E}
	E\equiv\frac{F(r)\dot{v}-\sqrt{F(r)G(r)}\dot{r}}{\sqrt{-F(r)\dot{v}^2+2\sqrt{F(r)G(r)}\dot{v}\dot{r}}}
\end{equation}
does not depend on $\lambda$, {\it i.e.}, is ``conserved". Note that $E$ does depend on the boundary time $t_{\rm R}$. Also, since \eqref{eq:volume} is invariant under the reparametrization of $\lambda$, we choose a gauge
\begin{equation}
\label{eq:proper volume gauge}
	\sqrt{-F(r)\dot{v}^2+2\sqrt{F(r)G(r)}\dot{v}\dot{r}} = 1\, .
\end{equation}
From \eqref{eq:E} and \eqref{eq:proper volume gauge}, we obtain
\begin{equation}
\label{eq:r dot and v dot over r dot}
	\dot{r}=\sqrt{\frac{F(r)+E^2}{F(r)G(r)}}\, , \quad \frac{\dot{v}}{\dot{r}}=\frac{\sqrt{F(r)G(r)}}{F(r)}\left( \frac{E}{\sqrt{F(r)+E^2}}+1 \right)\, .
\end{equation}
Recalling $r(0)=r_{\rm min}<r_{\rm h}$ and $\dot{r}(0)=0$, the first equation of \eqref{eq:r dot and v dot over r dot} with $\lambda=0$ yields
\begin{equation}
\label{eq:E^2}
	E^2=-F(r_{\rm min}) >0\, .
\end{equation}
As we will see shortly, $E$ must be negative. Substituting \eqref{eq:proper volume gauge} and \eqref{eq:r dot and v dot over r dot} into \eqref{eq:volume}, the extremal volume is written as
\begin{align}
	V
	&= 2\Omega_{k,d-1} \int_{r_{\rm min}}^{r_{\rm max}}\frac{dr}{\dot{r}} \notag \\
	&= 2\Omega_{k,d-1} \int_{r_{\rm min}}^{r_{\rm max}}dr~\sqrt{\frac{F(r)G(r)}{F(r)+E^2}}\, . \label{eq:extremal volume}
\end{align}
On the other hand, integrating $dt+dr^*=dv$ over $0\leq\lambda\leq\lambda_{\rm max}$ and using \eqref{eq:tortoise when f g indep} and \eqref{eq:r dot and v dot over r dot}, we obtain
\begin{align}
	t_{\rm R} - t(0) + \int_{r_{\rm min}}^{r_{\rm max}}dr~\frac{\sqrt{F(r)G(r)}}{F(r)}
	&= \int_{r_{\rm min}}^{r_{\rm max}}dr~\frac{\dot{v}}{\dot{r}} \notag \\
	&= \int_{r_{\rm min}}^{r_{\rm max}}dr~\frac{\sqrt{F(r)G(r)}}{F(r)}\left( \frac{E}{\sqrt{F(r)+E^2}}+1 \right)\, . \label{eq:old t_R}
\end{align}
Due to the symmetry of the spacetime, we find $t(0)=0$. Near the horizon, the integrand of the integral on the left-hand side behaves like
\begin{equation}
		\frac{\sqrt{F(r)G(r)}}{F(r)} \propto \frac{1}{r-r_{\rm h}}\, .
\end{equation}
Hereafter, we regard the integral as the Cauchy principal value, which is finite. Then, \eqref{eq:old t_R} simplifies to
\begin{equation}
\label{eq:t_R}
	t_{\rm R}=\int_{r_{\rm min}}^{r_{\rm max}}dr~\frac{\sqrt{F(r)G(r)}}{F(r)}\frac{E}{\sqrt{F(r)+E^2}}\, .
\end{equation}

It is natural to assume that $r_{\rm min}$ decreases as $t_{\rm R}$ increases since the wormhole should be more stretched in the time evolution. With \eqref{eq:t_R}, this implies that $E$ is negative, and \eqref{eq:E^2} yields
\begin{equation}
\label{eq:E in terms of F}
	E=-\sqrt{-F(r_{\rm min})}\, .
\end{equation}
Multiplying \eqref{eq:t_R} by $E$, we get
\begin{equation}
	Et_{\rm R}=\int_{r_{\rm min}}^{r_{\rm max}}dr~\frac{\sqrt{F(r)G(r)}}{F(r)}\sqrt{F(r)+E^2}-\frac{V}{2\Omega_{k,d-1}}\, .
\end{equation}
Taking the derivative with respect to $t_{\rm R}$, we obtain
\begin{equation}
\label{eq:dV/dt_R}
	\frac{dV}{dt_{\rm R}} = -2\Omega_{k,d-1}E = 2\Omega_{k,d-1}\sqrt{F(r_{\rm min})}\, .
\end{equation}
At late times, this asymptotically approaches the constant value
\begin{equation}
\label{eq:late time}
	\lim_{t_{\rm R}\to\infty}\frac{dV}{dt_{\rm R}}=2\Omega_{k,d-1}\sqrt{-F(r_{\rm f})}\, .
\end{equation}
Here, $r=r_{\rm f}$ is the radius where $\sqrt{-F(r)}$ is maximal. As $t_{\rm R}\to\infty$, the extremal surface asymptotically approaches this $r=r_{\rm f}$ surface.

\vspace{5mm}
\subsubsection{The Hartman-Maldacena entanglement entropy}

Presently, $f$ and $g$ are treated as being independent with each other. Therefore, reconstructing both of them requires as the inputs two functional degrees of freedom, one of which is given by the complexity. In the present paper, as an additional quantity as supplement, 
we use the Hartman-Maldacena entanglement entropy $S_{\rm HM}$ \cite{Hartman:2013qma}.

Holographically, $S_{\rm HM}$ is given by the volume $A$ of the Ryu-Takayanagi surface \cite{Ryu:2006bv} as
\begin{equation}
\label{eq:RT}
	S_{\rm HM}=\frac{A}{4G_{\rm N}}\, .
\end{equation}
Repeating the calculations in Sec.~\ref{subsec:volume}, we obtain
\begin{equation}
\label{eq:A}
	A=2\Omega_{k,d-2}\int_{\tilde{r}_{\rm min}}^{r_{\rm max}} dr~\sqrt{\frac{\tilde{F}(r)\tilde{G}(r)}{\tilde{F}(r)-\tilde{F}_{\rm min}}}\, ,
\end{equation}
\begin{equation}
\label{eq:HM t_R}
	t_{\rm R} = \int_{\tilde{r}_{\rm min}}^{r_{\rm max}} dr~\frac{\sqrt{\tilde{F}(r)\tilde{G}(r)}}{\tilde{F}(r)}\frac{-\sqrt{-\tilde{F}(\tilde{r}_{\rm min})}}{\sqrt{\tilde{F}(r)-\tilde{F}(\tilde{r}_{\rm min})}}\, , 
\end{equation}
where $\tilde{F}(r)\equiv f(r)r^{2(d-2)},\tilde{G}(r)\equiv g(r)r^{2(d-2)}$. 
Note that the only difference from the equations for the complexity in Sec.~\ref{sec:3.1.1} is the power of $r$
in the definitions of $\tilde{F}(r)$ and $\tilde{G}(r)$.
Resultantly, 
$\tilde{r}_{\rm min}$ may be different from $r_{\rm min}$ of the codimension-one extremal surface. 

Also, we can show
\begin{equation}
	\frac{dA}{dt_{\rm R}}=2\Omega_{k,d-2}\sqrt{-\tilde{F}(\tilde{r}_{\rm min})}\, .
\end{equation}
At late times, this asymptotically approaches the constant value
\begin{equation}
	\lim_{t_{\rm R}\to\infty}\frac{dA}{dt_{\rm R}}=2\Omega_{k,d-2}\sqrt{-\tilde{F}(\tilde{r}_{\rm f})}\, .
\end{equation}
Here, $r=\tilde{r}_{\rm f}$ is the radius at which $\sqrt{-\tilde{F}(r)}$ is maximal. As $t_{\rm R}\to\infty$, the extremal surface asymptotically approaches this $r=\tilde{r}_{\rm f}$ surface. 

We here have seen that all the equations for the Hartman-Maldacena entanglement entropy
are of the same form as those for the complexity of the CV conjecture. The only difference is in the power of $r$ in the integrands. For the reconstruction, 
we can use this slight difference to extract the
information of $f(r)$ and $g(r)$ independently.

\clearpage

\subsection{Reconstruction formula}
\label{subsec:reconstruction of f and g}

\begin{screen}
Assume that
\begin{itemize}
\item the time derivative of the complexity $\dot{\cal C}(t)$ and 
the time derivative of the Hartman-Maldacena entanglement entropy $\dot{S}_{\rm HM}(t)$ in the dual QFT are given. 
\\[-8mm]
\item $\dot{\cal C}(t)$ and $\dot{S}_{\rm HM}(t)$ are monotonic in time.
\\[-8mm]
\item the metric \eqref{eq:f g indep case} outside the horizon $(r \geq r_{\rm h})$ is known.
\end{itemize}
Then, solve
\begin{equation}
\label{eq:dC/dt_R and dS_HM/dt_R}
	\frac{d{\cal C}}{dt_{\rm R}}(t_{\rm R})=\frac{2\Omega_{k,d-1}}{G_{\rm N}L}\sqrt{-F_{\rm min}}\, , \quad \frac{dS_{\rm HM}}{d\tilde{t}_{\rm R}}(\tilde{t}_{\rm R})=\frac{\Omega_{k,d-2}}{2G_{\rm N}}\sqrt{-\tilde{F}_{\rm min}}
\end{equation}
to get $t_{\rm R}=t_{\rm R}(F_{\rm min})$ and $\tilde{t}_{\rm R}=\tilde{t}_{\rm R}(\tilde{F}_{\rm min})$. In addition, outside the horizon, invert $F=F(r)\equiv f(r)r^{2(d-1)}$ and $\tilde{F}=\tilde{F}(r)\equiv f(r)r^{2(d-2)}$ to find $r=r(F)$ and $r=\tilde{r}(\tilde{F})$. Substitute them into
\begin{equation}
\label{eq:sec 3 Q}
	Q(F_{\rm min}) \equiv \frac{t_{\rm R}(F_{\rm min})}{\sqrt{-F_{\rm min}}} + \int_{\epsilon}^\infty dF~\frac{dr}{dF}\frac{\sqrt{FG(r(F))}}{F}\frac{1}{\sqrt{F-F_{\rm min}}}\, ,
\end{equation}
\begin{equation}
\label{eq:sec 3 Q tilde}
	\tilde{Q}(\tilde{F}_{\rm min}) \equiv \frac{\tilde{t}_{\rm R}(\tilde{F}_{\rm min})}{\sqrt{-\tilde{F}_{\rm min}}} + \int_{\delta}^\infty d\tilde{F}~\frac{d\tilde{r}}{d\tilde{F}}\frac{\sqrt{\tilde{F}\tilde{G}(\tilde{r}(\tilde{F}))}}{\tilde{F}}\frac{1}{\sqrt{\tilde{F}-\tilde{F}_{\rm min}}}\, ,
\end{equation}
with positive infinitesimal $\epsilon$ and $\delta$, and $G(r) \equiv g(r)r^{2(d-1)}$ and $\tilde{G}(r)\equiv g(r)r^{2(d-2)}$. For $F\leq-\epsilon$ and $\tilde{F}\leq-\delta$, calculate
\begin{equation}
\label{eq:sec 3 y}
	y(F) \equiv \frac{1}{\pi}\frac{d}{dF}\int_F^{-\epsilon}dU~\frac{Q(U)}{\sqrt{U-F}}\, ,
\end{equation}
\begin{equation}
\label{eq:sec 3 y tilde}
	\tilde{y}(\tilde{F}) \equiv \frac{1}{\pi}\frac{d}{d\tilde{F}}\int_{\tilde{F}}^{-\delta}d\tilde{U}~\frac{\tilde{Q}(\tilde{U})}{\sqrt{\tilde{U}-\tilde{F}}}\, .
\end{equation}
Using these, solve
\begin{equation}
	\frac{d\tilde{F}}{dF}=\frac{\tilde{y}(\tilde{F})}{y(F)}
	\label{FFtilde}
\end{equation}
to find $\tilde{F}$ as a function of $F$. 
Substituting it into $r=F/\tilde{F}$, we obtain $r=r(F)$, from which $f(r)$ is obtained. 
Finally, substitute $r=r(F)$ into the equation
\begin{equation}
\label{eq:sec 3 diff eq}
	\frac{dr}{dF}\frac{\sqrt{FG(r(F))}}{F} = y(F)\, , 
\end{equation}
and solve it for $G(r(F))$. From that, $g(r)$ is obtained. This determines the metric inside
the horizon, $f(r)$ and $g(r)$ for $r<r_{\rm h}$.
\end{screen}

\begin{proof}
For the logic until \eqref{eq:sec 3 y tilde}, read the proof of Sec.~\ref{sec:2}.
Then the following differential equations are derived which the metric functions should satisfy, 
\begin{equation}
\label{eq:sec 3 diff eq 2}
	\frac{dr}{dF}\frac{\sqrt{FG(r(F))}}{F} = y(F)\, , \quad \frac{d\tilde{r}}{d\tilde{F}}\frac{\sqrt{\tilde{F}\tilde{G}(\tilde{r}(\tilde{F}))}}{\tilde{F}} = \tilde{y}(\tilde{F})\, .
\end{equation}
The first equation is the same as \eqref{eq:sec 3 diff eq}.
Note that $F$ and $\tilde{F}$ are not independent but constrained by $r(F)=\tilde{r}(\tilde{F})$. Using $F(r)=r\tilde{F}(r)$ and $G(r)=r\tilde{G}(r)$, we find that
\eqref{eq:sec 3 diff eq 2} yields \eqref{FFtilde}, which is the last nontrivial
equation to derive $F(r)$.
\end{proof}
%
%

\vspace{10mm}


\section{Example: BTZ black hole}
\label{sec:4}

In this section, we provide an explicit example to show that our reconstruction formulas do work properly.
The example is the static BTZ black hole,
\begin{equation}
\label{eq:BTZ}
	ds^2=-f(r)dt^2+\frac{dr^2}{f(r)}+r^2d\phi^2\, , \quad f(r)=\frac{r^2-r_{\rm h}^2}{L^2}\, .
\end{equation}
We employ the BTZ black hole because the analytic form of the Hartman-Maldacena entanglement entropy is known \cite{Hartman:2013qma}.
%
It is given by the volume of the codimension-two extremal surface linking the left and the right boundaries, which is just the geodesic length,
\begin{equation}
\label{eq:BTZ S_HM}
	S_{\rm HM}=\frac{A}{4G_{\rm N}}\, , \quad A=2L\left[ \ln\frac{2r_{\rm max}}{r_{\rm h}}+\ln\cosh\frac{r_{\rm h} t_{\rm R}}{L^2} \right]\, ,
\end{equation}
where $r_{\rm max}$ is the UV cutoff. Therefore, we have
\begin{equation}
\label{eq:BTZ growth rate}
	\frac{dA}{dt_{\rm R}}=2\frac{r_{\rm h}}{L}\tanh\frac{r_{\rm h} t_{\rm R}}{L^2}\, .
\end{equation}

Although we do not know the exact form of the complexity, it is obvious that 
the formulas we have obtained in Sec.~\ref{sec:2} for $g(r)=1/f(r)$ can be generalized 
also to the Hartman-Maldacena entanglement entropy, we can use 
this \eqref{eq:BTZ growth rate} as a test case to verify our reconstruction method.

Pretending that we do not know $f(r)$ for the interior of the black hole, we follow our reconstruction formula and obtain the metric of the form \eqref{eq:f metric} inside the horizon. We assume that we know \eqref{eq:BTZ growth rate} and also $f(r)$ outside the horizon.
We compare our obtained metric with the original analytic form \eqref{eq:BTZ}. If the original 
metric \eqref{eq:BTZ} is well reconstructed, we can say that our formulas work well.

\begin{figure}
	\centering
	\includegraphics[width=120mm]{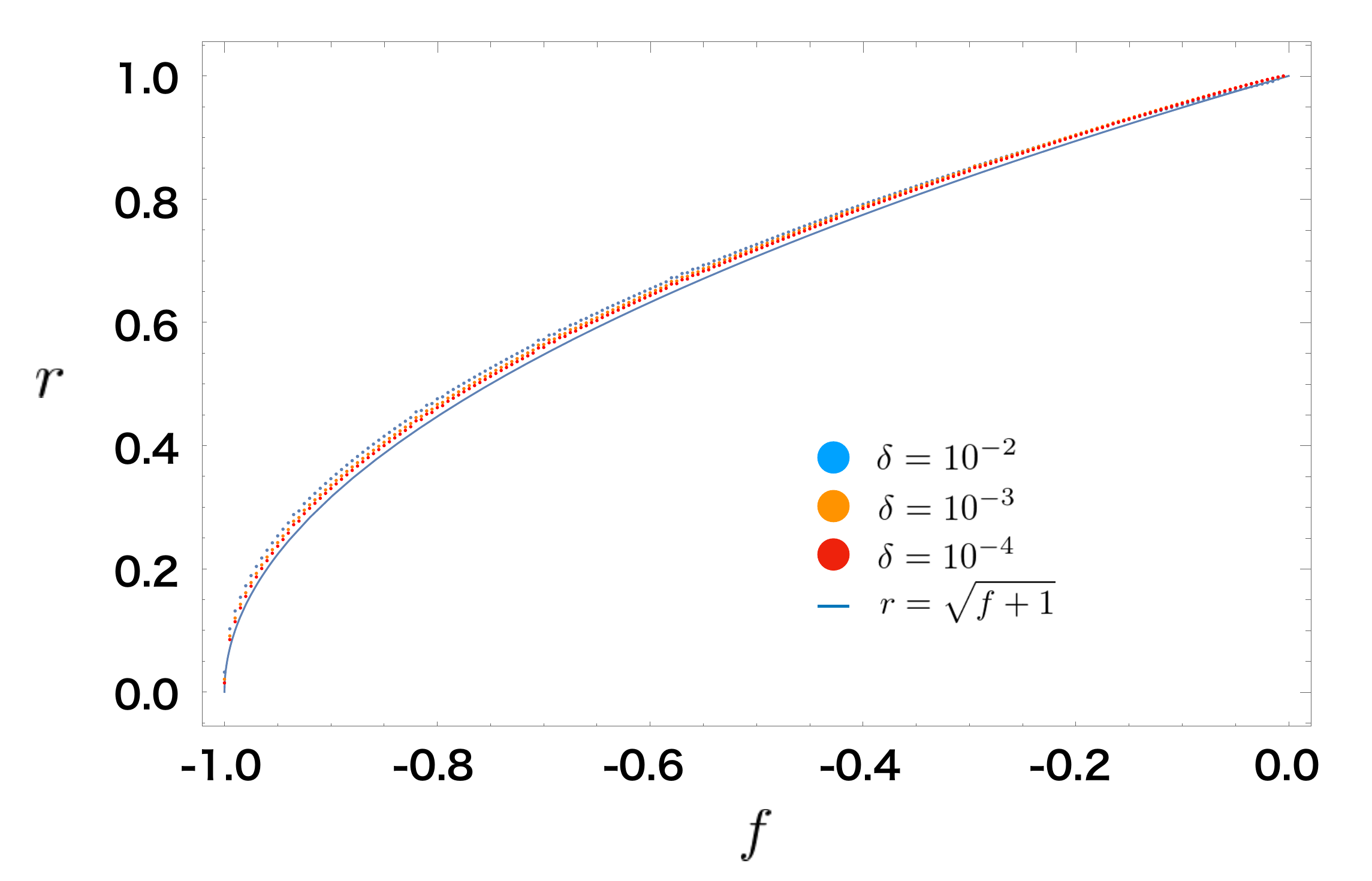}
	\caption{The numerical results of \eqref{eq:r=r(f)} for $\delta=10^{-2},10^{-3},10^{-4}$ (dotted lines, from top to bottom for each value of $\delta$ respectively), and the expected analytic form $r=\sqrt{f+1}$ (solid line). The metric of the interior of the BTZ black hole is properly reconstructed. We can see the reconstruction works more accurately for smaller $\delta$.}
	\label{fig:figure}
\end{figure}

For simplicity, we set $r_{\rm h}=L=1$ in what follows. Also, since $\tilde{F}=f$ for $d=2$, we simply write as $f$ instead of $\tilde{F}$. First, we solve\footnote{For $d=2$, $\Omega_{k,d-2}$ is unity.}
\begin{equation}
	\frac{dS_{\rm HM}}{dt_{\rm R}}=\frac{1}{2G_{\rm N}}\sqrt{-f_{\rm min}}\, , 
\end{equation}
to find $t_{\rm R}$ as a function of $f_{\rm min}$. The result is 
\begin{equation}
	t_{\rm R}=\tilde{t}_{\rm R}(f_{\rm min})\equiv {\rm Arctanh}\left(\sqrt{-f_{\rm min}}\right)\, .
\end{equation}
Next, using the metric $f(r)=r^2-1$ outside the horizon and $\tilde{G}=g=1/f$, we rewrite \eqref{eq:sec 3 Q tilde} as
\begin{align}
	\tilde{Q}(f_{\rm min})
	&= \frac{{\rm Arctanh}\left(\sqrt{-f_{\rm min}}\right)}{\sqrt{-f_{\rm min}}} + \int_\delta^\infty df~\frac{1}{f\sqrt{(f+1)(f-f_{\rm min})}} \notag \\
	&= \frac{{\rm Arctanh}\left(\sqrt{-f_{\rm min}}\right)}{\sqrt{-f_{\rm min}}} + \int_\xi^{\pi/2} d\theta~\frac{1}{\sin\theta \sqrt{\tan^2\theta - f_{\rm min}}}\, , \label{eq:BTZ Q tilde}
\end{align}
where we have set $f=\tan^2\theta$ and have defined $\xi\equiv {\rm Arctan}\sqrt{\delta}$ in the last line. We numerically calculate \eqref{eq:BTZ Q tilde} for a various choice of the infinitesimal parameter, $\delta=10^{-2},10^{-3},10^{-4}$. From \eqref{eq:sec 3 y tilde} and \eqref{eq:sec 3 diff eq 2}, we understand that the equation we have to solve is
\begin{equation}
	\frac{dr}{df}\frac{1}{f} = \frac{1}{\pi}\frac{d}{df}\int_f^{-\delta}du~\frac{\tilde{Q}(u)}{\sqrt{u-f}}\, ,
\end{equation}
which we rewrite as
\begin{equation}
\label{eq:derivative}
	dr = df \frac{1}{\pi}f\frac{d}{df}\int_f^{-\delta}du~\frac{\tilde{Q}(u)}{\sqrt{u-f}}\, .
\end{equation}
Near the horizon, $\delta$ is the small cutoff for the value of $f$. 
Correspondingly, the horizon cutoff for $r$ is found as  $r_\delta \equiv \sqrt{1-\delta}$
since we know the analytic structure of the metric near the horizon. 
Using this as a upper limit of the integral region, we integrate \eqref{eq:derivative},
\begin{equation}
	\int_r^{\sqrt{1-\delta}}dr=\int_f^{-\delta} df' \frac{1}{\pi}f'\frac{d}{df'}\int_{f'}^{-\delta}du~\frac{\tilde{Q}(u)}{\sqrt{u-f'}}\, .
\end{equation}
A partial integration yields
\begin{equation}
\label{eq:r=r(f)}
	r=\sqrt{1-\delta} + \frac{1}{\pi}\left( f\int_{f}^{-\delta}du~\frac{\tilde{Q}(u)}{\sqrt{u-f}}+ \int_f^{-\delta} df' \int_{f'}^{-\delta}du~\frac{\tilde{Q}(u)}{\sqrt{u-f'}} \right)\, .
\end{equation}
Numerically calculating the right-hand side of this equation, we finally obtain $r=r(f)$ inside the horizon. See Fig.\ref{fig:figure} for the plots of our numerical results. 
We can see that our reconstruction formulas neatly reproduce the interior of the BTZ black hole. The $\delta$ dependence is as expected --- the reconstruction works more accurately for smaller $\delta$. This BTZ case is special in that the final RT surface at $t_{\rm R}\to\infty$ 
hits the singularity, thus we are eventually able to reconstruct the whole interior of the BTZ black hole.

\clearpage

\section{Summary and discussions}
\label{sec:5}

In this paper, we have obtained formulas to reconstruct a metric inside a black hole horizon by the time derivative of the complexity measured in the dual QFT, based on the ``complexity = volume" (CV) conjecture in the holographic duality.
We have assumed only that the metrics outside the horizon is known. 

The bulk metric in the stationary spacetimes takes the form 
$ds^2 = -f(r)dt^2 + g(r) dr^2 + r^2 d\Omega^2$. If $g(r)$ is known as a function of $f(r)$
(such as the case $g(r)=1/f(r)$), 
we need only the time dependence of the complexity in the dual QFT as the data to reconstruct
the metric of the black hole interior, as we have seen in Sec.~\ref{subsec:reconstruction of f} 
and in App.~\ref{app:g(f) case}.
On the other hand, in the general case when both $f(r)$ and $g(r)$ are not known, 
we need, in addition to the complexity, 
the Hartman-Maldacena entanglement entropy of the doubled QFT \cite{Hartman:2013qma} 
to reconstruct the interior metric, as we have seen in Sec.~\ref{subsec:reconstruction of f and g}.

For our reconstruction formula, the metric outside the horizon needs to be known. We note here 
that other QFT quantities can determine it through the holographic duality. 
The entanglement entropy can determine only $g(r)$ \cite{Bilson:2008ab,Bilson:2010ff}
because the RT surface is on a time slice. The Wilson loop \cite{Hashimoto:2020mrx} can determine both
$f(r)$ and $g(r)$ but they are in the string frame, and they differ from the Einstein frame by
a conformal factor of unknown dilaton profile. The method using the light-cone cuts \cite{Engelhardt:2016wgb,Engelhardt:2016crc,Hernandez-Cuenca:2020ppu} determines the metric up to
a conformal factor. 
Therefore, the combination of the entanglement entropy and either 
the Wilson loops or the light-cone cuts
can determine the metric outside the black hole horizon completely.

Another assumption for the reconstruction of the metric
is that the time derivative of the complexity needs to be monotonic.
The charged and uncharged AdS black holes in various dimensions satisfy this
criterion \cite{Carmi:2017jqz}, while some field-theoretic definitions of complexity may not satisfy it \cite{Kim:2017qrq}. There exist several different proposals on the 
definition of the complexity in QFTs \cite{Susskind:2014jwa,Chapman:2017rqy,Caputa:2017urj,Hashimoto:2017fga,Yang:2017nfn,Yang:2018nda,Camargo:2019isp,Yang:2019udi}, and the time dependence can vary depending on the definitions
(see \cite{Carmi:2017jqz,Kim:2017qrq,Yang:2016awy,Brown:2017jil,Cottrell:2017ayj,Hashimoto:2018bmb,Ali:2018fcz} for the study of
the time dependence). 
We plan to come back to the question of what our formula can say for a holographically 
consistent definition of the QFT complexity.

How far can we go with this reconstruction method to reveal the mechanism of the AdS/CFT correspondence? Since the gravity side of the AdS/CFT are described mainly by geometric quantities,
once the inverse problems to obtain spacetimes by the probes
can be solved analytically as demonstrated in this paper,
enough information of the dual QFT data will be able to reconstruct completely 
the bulk metrics and the bulk theory. Using the complexity would be beneficial since it can go
even inside of the horizons, and our spirit is close to the derivation of the Einstein equations
by varying the complexity \cite{Czech:2017ryf}. The holographic complexity is a version of
surfaces in quantum gravity spacetimes, and string theory has been developed based
on probes of surfaces, which started with the brane democracy \cite{Townsend:1995gp}.
Needless to mention a similarity between the complexity surface and the spacelike branes \cite{Gutperle:2002ai,Hashimoto:2002sk}, the various surfaces in the bulk may keep playing 
a central role in the study of the AdS/CFT correspondence.

\acknowledgments

We are grateful to K.~Y.~Kim for discussions.
K.~H.~was supported in part by JSPS KAKENHI Grant No.~JP17H06462.

\clearpage

\appendix

\section{Treatment of the special case $g=g(f)$}
\label{app:g(f) case}

As we have seen in Sec.~\ref{sec:2}, if we assume $g=1/f$, we can reconstruct the metric inside the horizon by the time dependence of the complexity once the metric outside the horizon is known. More generally, 
if $g$ is known as a function of $f$, we can reconstruct the geometry by only the complexity.
Here we present our formula for that case.

\begin{screen}
Assume that
\begin{itemize}
\item the time derivative of the complexity $\dot{\cal C}(t)$ in the dual QFT is given. 
\\[-8mm]
\item $\dot{\cal C}(t)$ is monotonic in time.
\\[-8mm]
\item the metric \eqref{eq:f g indep case} outside the horizon $(r \geq r_{\rm h})$ is known.
\end{itemize}
Then, solve
\begin{equation}
	\frac{d{\cal C}}{dt_{\rm R}}(t_{\rm R})=\frac{2\Omega_{k,d-1}}{G_{\rm N}L}\sqrt{-F_{\rm min}}
\end{equation}
to get $t_{\rm R}$ as a function of $F_{\rm min}$. In addition, outside the horizon, invert $F=F(r)\equiv f(r)r^{2(d-1)}$ to find $r=r(F)$. Substitute them into
\begin{equation}
	Q(F_{\rm min}) \equiv \frac{t_{\rm R}(F_{\rm min})}{\sqrt{-F_{\rm min}}} + \int_{\epsilon}^\infty dF~\frac{dr}{dF}\frac{\sqrt{FG(F,r(F))}}{F}\frac{1}{\sqrt{F-F_{\rm min}}}\, ,
\end{equation}
with a positive infinitesimal parameter $\epsilon$. $G(F,r(F))$ is defined as
\begin{equation}
	G(F,r(F)) \equiv g(Fr(F)^{-2(d-1)})r(F)^{2(d-1)}\, .
\end{equation}
For $F\leq-\epsilon$, calculate
\begin{equation}
	y(F) \equiv \frac{1}{\pi}\frac{d}{dF}\int_F^{-\epsilon}dU~\frac{Q(U)}{\sqrt{U-F}}\, ,
\end{equation}
and integrate the differential equation
\begin{equation}
	\frac{dr}{dF}\frac{\sqrt{FG(F,r(F))}}{F} = y(F)
\end{equation}
to find $r=r(F)$. Finally, invert it to find $F(r)$, and the metric inside the black hole is given by $f(r)=F(r)r^{-2(d-1)}$.
\end{screen}

The proof is almost identical to that of Sec.~\ref{subsec:reconstruction of f} since \eqref{eq:t_R} and \eqref{eq:dV/dt_R} have the same form as \eqref{eq:sec 2 t_R} and \eqref{eq:sec 2 dV/dt_R}, respectively.

\clearpage

\end{document}